\documentclass{iopart}

\usepackage{graphicx}
\usepackage{cite}
\newcommand{\bea}{\begin{eqnarray}}
\newcommand{\eea}{\end{eqnarray}}
\newcommand{\be}{\begin{equation}}
\newcommand{\ee}{\end{equation}}
\renewcommand{\Im}{\mathop{\mathrm{Im}}}

\begin{document}

\title[Spin correlations in spin blockade]{Spin correlations in spin blockade}

\author{Rafael S\'anchez$^{1}$, Sigmund Kohler$^2$ and Gloria Platero$^1$}

\address{$^1$ Instituto de Ciencia de Materiales de Madrid
(CSIC), Cantoblanco, 28049 Madrid, Spain}
\address{$^2$ Institut f\"ur Physik, Universit\"at Augsburg, Universit\"atstra\ss e 1, 86135 Augsburg, Germany}
\ead{gloria.platero@icmm.csic.es}

\begin{abstract}
We investigate spin currents and spin-current correlations for double
quantum dots in the spin blockade regime. By analysing the time
evolution of the density matrix, we obtain the spin resolved currents
and derive from a generating function an expression for the
fluctuations and correlations. Both the charge current and the spin
current turn out to be generally super-Poissonian. Moreover, we study
the influence of ac fields acting upon the transported electrons.  In
particular, we focus on fields that cause spin rotation or
photon-assisted tunnelling.
\end{abstract}


\section{Introduction}

The recently achieved access to individual electron states of quantum
dots \cite{hansonrev} spurred the interest in the control of single
electron degrees of freedom in nanoconductors providing a designable
alternative to atoms in quantum optics.  In particular by applying ac
fields, interesting effects such as coherent destruction of electron
tunnelling by electric ac fields \cite{grossman,grossman2,pr,sigmundRep,
dellaValle,to} or electron spin rotations by crossed magnetic ac and dc
fields \cite{engel,koppens,esr} have been reported.
A double quantum dot with up to two electrons provides a
perfect framework for spin manipulation via \textit{spin blockade}
\cite{weinmann,ono,inarrea,fransson}.  These systems are designed
such that the only state located in the transport window is the
doubly occupied singlet state. In this particular configuration,
for electrons occupying inter-dot triplet states, i.e.\ two-electron
states in which the electrons have the same spin orientation, inter-dot
tunnelling is suppressed due to the Pauli exclusion principle.
Tunnelling current in ac electric fields
\cite{pr,nanotech,spinpump,creffield} can be tuned by means
of the frequency and intensity of the field, which yields
interesting features like for instance charge localisation
within the quantum dot structure (dynamical localisation) and
suspension of Coulomb and spin blockade.
\begin{figure}[t]
\begin{center}
\includegraphics[width=2.95in,clip]{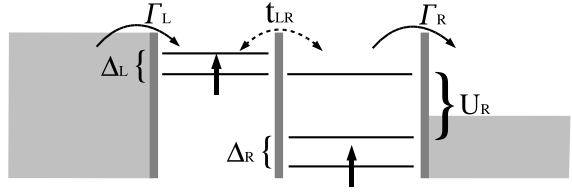}
\end{center}
\caption{\label{sb} Double quantum dot in spin blockade
configuration. The chemical potentials in the contacts are adjusted
such that the doubly occupied intra-dot singlet state is the only one
located in the transport window. As electrons fall in the inter-dot
triplet subspace, inter-dot tunnelling is forbidden by spin
conservation.  Consequently, transport is suppressed. }
\end{figure}

Spin blockade also plays an important role for coherent spin
manipulation, as demonstrated by electron spin resonance
experiments where a resonant ac magnetic field---perpendicular to
a dc one---acts on the double quantum dot \cite{koppens} giving
rise to coherent single-spin rotations. Under certain
configurations, the singlet states are populated such
that spin blockade is lifted and, thus, an electronic current
flows through the system. Then the current exhibits signatures of
an interplay between coherent spin rotations due to the ac
magnetic field and coherent electron delocalisation due to
interdot tunnelling \cite{esr}.

In this paper, we analyse the spin current and the spin current
fluctuations, as well as their correlations in double quantum dots
in the spin blockade regime in the presence of ac electric and
magnetic fields.  We demonstrate that spin current correlations
provide information about the spin dynamics.
Moreover, we also consider the case of spin relaxation which stems
from spin flip processes accompanied by the coupling to a bosonic heat bath.  This heat bath together
with the fermionic electron reservoirs forms an unconventional
environment for the quantum dots.

\section{Model and method}
\label{model}

We consider two weakly coupled quantum dots connected to two fermionic
leads (cf. figure \ref{sb}) described by the Hamiltonian
\begin{equation}
\label{h0}
\hat H=\hat H_0+\hat{H}_{\rm LR}+\hat H_{\rm T}+\hat H_{\rm leads},
\end{equation}
where $\hat{H}_0=\sum_{i\sigma}\varepsilon_{i\sigma}\hat
c_{i\sigma}^\dagger\hat c_{i\sigma}+\sum_{i}U_{i}\hat
n_{i{{\uparrow}}}\hat n_{i{\downarrow}}+U_{\rm LR}\hat n_{\rm
L}\hat n_{\rm R}$ describes the two isolated quantum dots,
$\hat{H}_{\rm LR}=-\sum_{\sigma}(t_{\rm LR}\hat{c}_{\rm
L\sigma}^\dagger\hat{c}_{\rm R\sigma}+{\rm H.c.})$ is the
inter-dot coupling.  $\hat H_{\rm T}=\sum_{l\epsilon\{\rm
L,R\}k\sigma}(\gamma_{l}\hat{d}_{lk\sigma}^\dagger\hat{c}_{l\sigma}+{\rm
H.c.})$ represents the tunnelling between the double quantum dot and the leads
described by $\hat{H}_{\rm leads}=\sum_{lk\sigma}\varepsilon_{lk\sigma}\hat
d_{lk\sigma}^\dagger\hat d_{lk\sigma}$. $\varepsilon_{i\sigma}$ is the
energy of an electron located in dot $i$ with spin $\sigma$, and $U_i$
($U_{\rm LR}$) is the intra-dot (inter-dot) Coulomb repulsion. 
In order to allow for up to two electrons in
the system (one in each dot), the chemical potentials of the leads,
$\mu_i$, must satisfy the condition $\varepsilon_i<\mu_i-U_{\rm
LR}<\varepsilon_i+U_i$ and $\mu_i<\varepsilon_i+2U_{\rm LR}$. In
addition, a dc magnetic field is introduced along the $z$-axis in
order to break the spin degeneracy by a Zeeman splitting
$\Delta_i=g_iB_{z,i}$, i.e.\ $\hat{H}_{\rm
B}=\sum_i\Delta_i\hat{S}_z^i$, where ${\bf
S}_i=\frac{1}{2}\sum_{\sigma \sigma'} c^\dagger_{i\sigma} {\bf
\sigma}_{\sigma \sigma'} c_{i\sigma'}$ are the spin operators of each
dot.

In this configuration, spin blockade is manifest for sufficiently
small bias voltage such that the state with two electrons in the right
dot (the state that supports the current) is in resonance with the one
electron states in each dot. The current is then quenched whenever the
electrons in each quantum dot have the same spin polarisation and the
Pauli exclusion principle avoids inter-dot tunnelling \cite{ono}.

Moreover, considering that any excited state in the right quantum dot has
an energy above the transport window and that inter-dot tunnelling is spin
independent, a current can flow only through processes
involving the doubly occupied singlet in the right quantum dot, $|S_{\rm
R}\rangle$ and the inter-dot singlet $|S_0\rangle
=(|{\uparrow},{\downarrow}\rangle
-|{\downarrow},{\uparrow}\rangle)/\sqrt{2}$.
Thus, the occupation of any inter-dot triplet state,
$|+\rangle=|{\uparrow},{\uparrow}\rangle$,
$|-\rangle=|{\downarrow},{\downarrow}\rangle$ and
$|T_0\rangle=(|{\uparrow},{\downarrow}\rangle+|{\downarrow},{\uparrow}\rangle)
/\sqrt{2}$
inhibits the transport to the collector \cite{esr}, unless, as we
will discuss below, the singlet and the triplet subspace mix due
to any perturbation, as for instance an inhomogeneous magnetic
field in the sample, which produces different Zeeman splittings
within each quantum dot.

\subsection{Master equation and full counting statistics}
The non-equilibrium dynamics of a quantum dot system can be
described by means of the equation of motion for the reduced
density operator $\rho$, obtained after tracing out the reservoirs in
the total density operator $R$:
\begin{equation}
\dot\rho= \tr_\mathcal{R}\dot R
= -\frac{i}{\hbar}\tr_{\cal R}[H,R]={\cal L}\rho,
\end{equation}
where $\mathcal{L}$ is the Liouvillian acting on the reduced density
operator. In matrix notation, it reads $\dot\rho_{i}={\cal
M}_{ij}\rho_j$, where $\rho_j$ denotes the density operator in vector
notation.

Accordingly, one can define current super-operators $\cal
J_\pm$ which, when acting on the reduced density operator,
describe the electron tunnelling from the quantum dots to the
collector and back, thus yielding positive and negative
contributions to the current. Then, the current can be formulated
as the trace of the current operator \cite{christianNEMS,franzSigmund}
\begin{equation}
I=e\tr_{\cal S}({\cal J}\rho)=e\tr_{\cal S}[({\cal J}_+-{\cal J}_-)\rho].
\end{equation}
Note that, in the same way as the super-operator $\cal L$ can be
written as a matrix and $\rho$ as a vector, the trace over the
system space can be expressed as multiplication with a transposed vector in
Liouvillian space, $v_0^\dagger$, which is the unit matrix in vector
notation.  Thus, the trace condition $\tr_{\cal S}\rho=1$ reads
$
v_0^\dagger\rho=1,
$
while trace conservation $\tr_{\cal S}\dot\rho = 0$ corresponds to the
relation
\begin{equation}
\label{v0+L}
v_0^\dagger{\cal L}=0
\end{equation}
and the current expectation value becomes
$I=ev_0^\dagger{\cal J}\rho$.

It is convenient to write the master equation with the help of the
current operators, which is easily done by identifying those
terms that change the number of particles in the collector
included in $\cal J_\pm$.  Then the master equation assumes the
form
\begin{equation}
\dot\rho={\cal L}(t)\rho=\left({\cal L}_0(t)+{\cal J}_++{\cal
J}_-\right)\rho.
\end{equation}
In a quantum dot system, the super-operator $\mathcal{L}_0(t)$
describes both the electron dynamics and tunnelling through the emitter
barrier.

Both current and shot noise can be expressed in terms of the accumulated
charge in the collector, $eN(t)$, such that \cite{blanter}
\begin{eqnarray}
\label{defI}
I &=& e\frac{\rmd}{\rmd t}\langle N(t)\rangle,
\\
\label{defS}
S &=& e^2\frac{\rmd}{\rmd t}\left(\langle N^2(t)\rangle-\langle N(t)\rangle^2\right).
\end{eqnarray}
Accordingly, one can obtain expressions for the higher order moments
of the current by evaluating the expectation value $\langle
N^\alpha\rangle$ that define the statistics of the transmitted
electrons---the \textit{full counting statistics}
\cite{levitovLesovik,bagretsNazarov}.

For the specific computation of the cumulants, we define the operator
\begin{equation}
\label{opG}
G(z,t)=\tr_{\cal R}\left(z^NR(t)\right) ,
\end{equation}
which generalises the density operator.  The latter is recovered in
the limit $G(z\to1,t)=\tr_{\cal R}R(t)=\rho(t)$.
$G(z,t)$ obeys the equation of motion
\begin{equation}
\label{meG}
\dot G(z,t)
=\left({\cal L}+(z-1){\cal J_+}+\left(z^{-1}-1\right){\cal J_-}\right)G(z,t) ,
\end{equation}
while its trace is the moment generating function for the
transported electrons, which means
\begin{equation}
\label{Nalfa}
\langle N^{\alpha}\rangle
=\left.v_0^\dagger\left(z\frac{\partial}{\partial z}\right)^\alpha G(z,t)\right|_{z=1}
\equiv \tr_{\cal S} g^{(\alpha)}(t),
\end{equation}
where
\begin{equation}
g^{(\alpha)}(t)
= \left.\left(z\frac{\partial}{\partial z}\right)^\alpha G(z,t)\right|_{z=1}
= \tr_{\cal R}\left(N^\alpha R(t)\right).
\end{equation}
Introducing the shorthand notation $g\equiv g^{(0)}$ and $g'\equiv
g^{(1)}$, we find the equations of motion \cite{franzSigmund}
\begin{eqnarray}
\label{G0}
\dot g(t) &=& \dot \rho(t)={\cal L}\rho(t)
\\
\label{G1}
\dot g'(t) &=& \mathcal{L}g'(t)+(\mathcal{J}_+-{\cal J_-})\rho(t) .
\end{eqnarray}
The corresponding equations of motion for the higher-order moments
allow the recursive computation of the full counting statistics
without the need of explicitly computing eigenfunctions and
derivatives \cite{flindtFCS}.  This is particularly helpful for larger
systems that require a numerical treatment.

\subsection{Stationary solution and zero frequency noise}
\label{sec:chargecurrent}
\begin{figure}[t]
\begin{center}
\includegraphics[width=2.95in,clip]{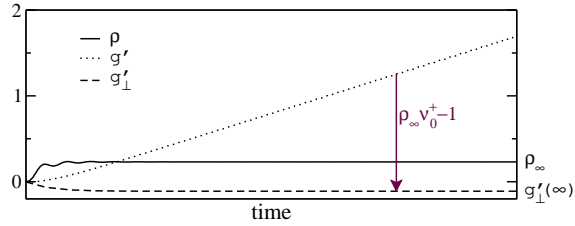}
\end{center}
\caption{\label{rhoges} Time evolution of a typical
diagonal element of the density matrix $\rho$ and the
corresponding terms of $g'$ and $g'_\perp$, see
section~\ref{sec:chargecurrent}. }
\end{figure}

The average current and the shot noise can be obtained by
integrating the equations of motion (\ref{G0}) and (\ref{G1}). For long
times, the system evolves into a stationary state determined by ${\cal
L}\rho_\infty=0$. The stationary solution $\rho_\infty$ thus
is the eigenvector of the Liouvillian that corresponds to the
eigenvalue zero with the normalisation $v_0^\dagger\rho_\infty=1$. By
inserting this into equation~(\ref{G1}), we obtain
\begin{equation}
\label{G1st} \dot g'(t)={\cal
L}g'(t)+({\cal J_+}-{\cal J_-})\rho_\infty.
\end{equation}
The eigenvalue zero of $\cal L$, which corresponds to the stationary
state, involves a solution with one component linear in time, which
can be singled out by projection to the null-space, $\rho_\infty
v_0^\dagger$, such that
\begin{equation}
\label{g'perp}
g'(t)=\rho_\infty
v_0^\dagger({\cal J_+}-{\cal J_-})\rho_\infty t+g'_\perp(t).
\end{equation}
The orthogonal component $g'_\perp(t)$, converges for long times, as
we will demonstrate below.  The information about the shot noise is
fully contained in $g'_\perp(t)$, and for its stationary solution, we
obtain by inserting (\ref{g'perp}) into (\ref{G1}) the algebraic
equation
\begin{equation}
\label{Lgperpinf}
{\cal L}g'_\perp(\infty)=(\rho_\infty v_0^\dagger-1)({\cal
J_+}-{\cal J_-})\rho_\infty,
\end{equation}
together with the orthogonality condition $v_0^\dagger
g'_\perp(\infty)=0$.  The numerical integration of equations~(\ref{G0}) and
(\ref{G1}) as well as the projection to the orthogonal subspace are
illustrated in figure~\ref{rhoges}.

Knowing $\rho_\infty$ and $g'_\perp(\infty)$ allows one to compute the
stationary current and the zero-frequency noise
\begin{eqnarray}
\label{in}
I &=& e ~v_0^\dagger({\cal J_+}-{\cal J_-})\rho_\infty
\\
\label{sn}
S &=& e^2v_0^\dagger \{ ({\cal J_+}-{\cal J_-})g'_\perp(\infty)
+({\cal J_+}+{\cal J_-})\rho_\infty \} ,
\end{eqnarray}
respectively.  The ratio between them defines the Fano factor
\begin{equation}
F=\frac{S}{e|I|} ,
\end{equation}
which reflects the sub- or super-Poissonian character of the noise
according to $F<1$ or $F>1$, respectively.  
This method is also valid
for ac driven systems provided that one consideres the averages $\bar
I$ and $\bar S$ over one period of the field \cite{franzSigmund}.

\subsection{Spin current fluctuations}

Up to now we considered the statistics of the {\it total} transferred
charge. However, we are also interested in the spin degree of freedom
an the eventual spin-spin correlations. It is straightforward to adapt
the formalism presented above to the case of transport of electrons
with spin $\sigma={\uparrow},{\downarrow}$ by replacing in
equations~(\ref{defI}) and (\ref{defS}) the electron number operator,
$N$, by number operator for electrons with spin $\sigma
={\uparrow},{\downarrow}$, $N_\sigma$. Then one obtains the spin
current operators ${\cal J}_{\sigma\pm}$ and the corresponding spin
currents
$I_\sigma=e\frac{\rmd}{\rmd t}\langle N_\sigma\rangle$, as well as the spin
auto-correlation $S_\sigma=e^2\frac{\rmd}{\rmd t}(\langle
N^2_\sigma\rangle-\langle N_\sigma\rangle^2)$ with the
spin Fano factor $F_\sigma=S_\sigma/(e|I_\sigma|)$.

Moreover, by comparing charge noise and spin noise, one can also study
cross-correlations of currents with opposite spin whose zero-frequency
component reads
\begin{equation}
S_{\uparrow\downarrow}
=\frac{\rmd }{\rmd t}(\langle N_\uparrow N_{\downarrow}\rangle
 -\langle N_\uparrow\rangle\langle N_{\downarrow}\rangle)
=\frac{1}{2}(S-\sum_{\sigma={\uparrow},{\downarrow}} S_\sigma).
\end{equation}
This quantity allows one to define the dimensionless spin
correlation coefficient
\begin{equation}
r=\frac{S_{\uparrow\downarrow}}{\sqrt{S_{\uparrow}S_{\downarrow}}}
\end{equation}
which assumes the value $r=1$ if the transport of
${\uparrow}$-electrons and ${\downarrow}$-electrons is perfectly
correlated, while $r=0$ for uncorrelated spin currents. 

\section{Spin current noise for undriven quantum dots}
\label{undriven}

In the configuration described in section~\ref{model}, i.e.\ for two
quantum dots with up to two electrons in the system,
it is convenient to decompose the density operator into the basis
$|1\rangle=|0,{\uparrow}\rangle$, $|2\rangle=|0,{\downarrow}\rangle$,
$|a\rangle=|{\uparrow},{\downarrow}\rangle$,
$|b\rangle=|{\downarrow},{\uparrow}\rangle$, $|S_{\rm
R}\rangle=|0,{\uparrow}{\downarrow}\rangle$,
$|+\rangle=|{\uparrow},{\uparrow}\rangle$,
$|-\rangle=|{\downarrow},{\downarrow}\rangle$.
As discussed above, transport suffers from spin blockade unless
the inter-dot triplet states have finite lifetime.  However, as
discussed in \cite{belzig}, transport becomes possible
for a small but finite thermal smearing of the Fermi surface of the
emitter.  Then an electron in the left quantum dot being in one of the states
$|a\rangle$, $|b\rangle$, $|\pm\rangle$ can tunnel with a rate
$x\Gamma_{\rm L}$ to the emitter, where $x=1-f(\varepsilon_{\rm
L}+U_{\rm LR})$, and can be replaced by an electron with opposite
spin. We neglect here the contribution of leakage currents due to
co-tunnelling \cite{etosb}.

Under the usual Born-Markov approximation \cite{blum}, the master
equation for the reduced density matrix, $\dot\rho={\mathcal L}\rho$,
in presence of an homogeneous magnetic field, i.e., for $\Delta_{\rm
L}=\Delta_{\rm R}$ reads
\begin{eqnarray}
\dot\rho_1&=&\Gamma_{\rm R}\rho_{S_{\rm R}}+x\Gamma_{\rm L}(\rho_b+\rho_+)-2(1-x)\rho_1\nonumber\\
\dot\rho_2&=&\Gamma_{\rm R}\rho_{S_{\rm R}}+x\Gamma_{\rm L}(\rho_a+\rho_-)-2(1-x)\rho_2\nonumber\\
\dot\rho_a&=&-2t_{\rm LR}\Im\rho_{S_{\rm R}a}-x\Gamma_{\rm L}\rho_a+(1-x)\Gamma_{\rm L}\rho_2\nonumber\\
\dot\rho_b&=&2t_{\rm LR}\Im\rho_{S_{\rm R}b}-x\Gamma_{\rm L}\rho_b+(1-x)\Gamma_{\rm L}\rho_1\\
\dot\rho_{S_{\rm R}}&=&2t_{\rm LR}\Im(\rho_{S_{\rm R}a}-\rho_{S_{\rm R}b})-2\Gamma_{\rm R}\rho_{S_{\rm R}}\nonumber\\
\dot\rho_+&=&(1-x)\Gamma_{\rm L}\rho_{1}-x\Gamma_{\rm L}\rho_+\nonumber\\
\dot\rho_-&=&(1-x)\Gamma_{\rm L}\rho_{2}-x\Gamma_{\rm L}\rho_-\nonumber
\end{eqnarray}
for the diagonal elements which describe occupation probabilities and
\begin{eqnarray}
\dot\rho_{ab}
&=& \rmi t_{\rm LR}(\rho_{S_{\rm R}b}+\rho_{aS_{\rm R}})
    -x\Gamma_{\rm L}\rho_{ab}\nonumber\\
\dot\rho_{aS_{\rm R}}
&=& \rmi t_{\rm LR}(\rho_{S_{\rm R}}-\rho_{a}+\rho_{ab})
   -\Big(\rmi\varepsilon+\frac{1}{2}(x\Gamma_{\rm L}
        +2\Gamma_{\rm R})\Big)\rho_{aS_{\rm R}}\\
\dot\rho_{bS_{\rm R}}
&=&- \rmi t_{\rm LR}(\rho_{S_{\rm R}}-\rho_{b}+\rho_{ba})
   -\Big(\rmi\varepsilon+\frac{1}{2}(x\Gamma_{\rm L}
   +2\Gamma_{\rm R})\Big)\rho_{bS_{\rm R}}\nonumber
\end{eqnarray}
for the off-diagonal elements, i.e.\ the coherences, where
$\varepsilon=\varepsilon_{\rm L}-\varepsilon_{\rm R}+U_{\rm LR}-U_{\rm
R}$.

The spin current operators $\cal J_{\sigma\pm}$ can be written as
matrices for which all elements but $({\cal
J}_{\uparrow+})_{2,S_{\rm R}}=({\cal J}_{\downarrow+})_{1,S_{\rm
R}}=\Gamma_{\rm R}$ vanish. Processes which transport an electron
from the collector to the state $|S_{\rm R}\rangle$ are
energetically forbidden and, thus, ${\cal J}_{\sigma-}=0$.

Under the condition $\varepsilon=0$, we find that $|a\rangle$ and
$|b\rangle$ are in resonance with $|S_{\rm R}\rangle$ such that for
symmetric dot-lead coupling $\Gamma_{\rm L}=\Gamma_{\rm R}=\Gamma$,
a leakage current $I=xe\Gamma/3$ emerges.  Then, the Fano factor
is determined by the internal dynamics even in the limit of small
thermal excitation $x\ll 1$ and reads
\begin{equation}
F=\frac{5}{3}-\left(\frac{7}{3}+\frac{4\Gamma^2}{9t_{\rm
LR}}\right)x+O(x^2).
\end{equation}
This super-Poissonian shot noise can be explained in terms of
electron bunching: Spin blockade occurs once electrons decay to
a triplet state. Then, the system will remain for a rather long
period in the triplet subspace, intermitted only by short lapses of
time in which an electron tunnels to the emitter and is replaced
by an electron with opposite spin and, thereby, resolves spin
blockade.  The resulting electron bunching turns into
super-Poissonian statistics, as in the case of dynamical channel
blockade \cite{belzig,prl,twochannels}, where two interacting
channels block each other.  Here the two channels are replaced by
the two interacting spin degrees of freedom.
\begin{figure}[t]
\begin{center}
\includegraphics[width=2.95in,clip]{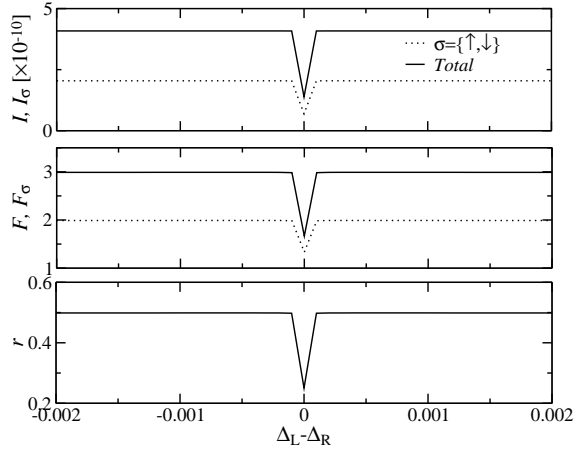}
\end{center}
\caption{\label{ifrEst} Current (in $\mathrm{mV}/\hbar$), Fano factors and spin
correlation for the undriven case, as a function of Zeeman
inhomogeneity for $t_{\rm LR}=0.0026$, $\varepsilon_{\rm L}=1.5$,
$\varepsilon_{\rm R}=0.45$, $U_{\rm R}=1.45$, $\Delta_{\rm
R}=0.026$ (corresponding to a magnetic field of $\sim 1{\rm T}$), $\mu_{\rm L}=1.94$, $\mu_{\rm R}=1.1$, $\Gamma_{\rm
L}=\Gamma_{\rm R}=10^{-3}$ and $k_{\rm B}T=0.001$ (energies in meV).}
\end{figure}

Interestingly enough, this result can be compared to the transport
through a single quantum dot for which one level (the singlet channel)
lies within the transport window and three levels (the three inter-dot
triplets) are below both Fermi surfaces, such that the occupation of
the later states induces dynamical channel blockade.  Then the Fano
factor becomes $F=1+(4/3)\Gamma_{\rm L}/(\Gamma_{\rm L}+\Gamma_{\rm
R})$ and for $\Gamma_{\rm L}=\Gamma_{\rm R}$ assumes the value
$F=5/3$.

In the absence of spin scattering processes, both spins contribute
equally to the current, $I_\sigma=I/2$, and the Fano factor for
each spin current reads
\begin{equation}
F_\sigma=\frac{4}{3}-\left(\frac{7}{6}+\frac{2\Gamma^2}{9t_{\rm LR}^2}\right)x+O(x^2).
\end{equation}
Moreover, we obtain the spin correlation coefficient
\begin{equation}
r=\frac{1}{4}-\frac{42\Gamma^2+21t_{\rm LR}^2}{32t_{\rm LR}^2}x+O(x^2).
\end{equation}

As can be seen in figure~\ref{ifrEst}, this behaviour considerably
changes as soon as an inhomogeneity in the Zeeman splittings appears.
Then mixing between $|S_0\rangle$ and $|T_0\rangle$ adds a new
conducting channel while the number of blocking states is reduced to
the two states $|+\rangle$ and $|-\rangle$ and, thus, the effective
transmitted charge and the Fano factor increase.  Then we find $F=3$,
$F_\sigma=2$, and $r=1/2$.

\section{Electron spin resonance}
\label{esr}

The spin of an electron can be manipulated by external magnetic
fields.  For instance, in the presence of an oscillating magnetic
field $B_{\rm ac}$ whose frequency equals the Zeeman splitting
produced by a constant magnetic field, the electron spins rotate,
which is know as electron spin resonance \cite{cohen}.
For quantum dots, this effect may manifest itself in current
oscillations \cite{engel} and allows one to determine the
spin scattering times \cite{dong}.

In a double quantum dot, one expects that the rotation of a single
electron spin removes spin blockade\cite{koppens}, but the presence of
a second electron can lead to collective rotations which quench the
current or affect the current oscillations\cite{esr}.  This behaviour
is reflected also in the shot noise characteristics, as we will find
below.

\begin{figure}[t]
\begin{center}
\includegraphics[width=2.95in,clip]{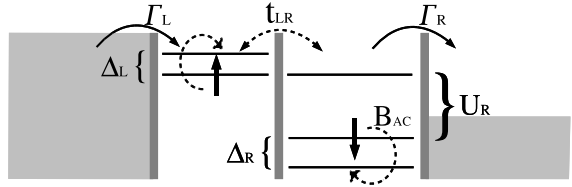}
\end{center}
\caption{\label{esqesr} Spin rotation in a double quantum dot
induced by an oscillating magnetic field at resonance with the Zeeman
splitting.
}
\end{figure}%
When a magnetic field with an ac component circularly polarised
perpendicular to the $x$-$y$-plane acts, then the $z$-components of
the electron spins start to rotate, as depicted
figure~\ref{esqesr}. The corresponding time-dependent Hamiltonian
reads
\begin{equation}
\hat{H}_{\rm B}(t)
=\sum_i\left[\Delta_i\hat{S}_z^i+gB_{\rm ac}\left(\hat{S}_x^i\cos\omega t+\hat{S}_y^i\sin\omega t\right)\right].
\end{equation}

In order to rotate the electron spin, the frequency of the ac magnetic
field must satisfy the resonance condition $\hbar\omega=\Delta_i$
\cite{engel}. Then one would naively expect that the Pauli exclusion
principle eventually would not apply to the interdot tunnelling such
that spatial oscillations between the left and right quantum dot
occurred and a finite current emerged.  For a homogeneous Zeeman
splitting through the double dot, $\Delta_{\rm L}=\Delta_{\rm R}$,
however, this is not the case. As the oscillating magnetic
field brings different spin projections within one dot into
resonance, the electron spins in different quantum dots rotate
simultaneously within the triplet subspace, which is decoupled from the
singlets. Then, the electrons become eventually trapped in the
triplet subspace and the tunnelling current towards the collector drops
to zero\cite{esr}.

\begin{figure}[t]
\begin{center}
\includegraphics[width=2.95in,clip]{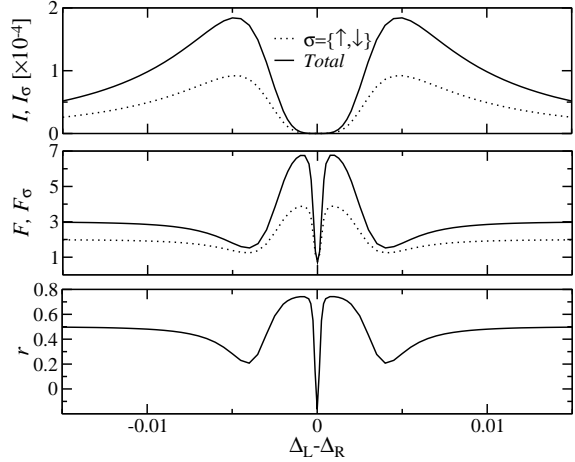}
\end{center}
\caption{\label{ifrBac} Current (in $\mathrm{mV}/\hbar$), Fano factors, and spin
correlation as a function of the Zeeman inhomogeneity for magnetic
field intensity $B_{\rm ac}=0.1\sqrt{2}{\rm T}$, $t_{\rm LR}=0.0026$,
$\varepsilon_{\rm L}=1.5$, $\varepsilon_{\rm R}=0.45$, $U_{\rm
R}=1.45$, $\hbar\omega=\Delta_{\rm R}=0.026$, $\mu_{\rm L}=2$, $\mu_{\rm
R}=1.1$, $\Gamma_{\rm L}=\Gamma_{\rm R}=10^{-3}$, $g=0.026$ (energies in
meV). }
\end{figure}

The Zeeman splittings in different dots may as well be different,
$\Delta_{\rm L}\neq\Delta_{\rm R}$,
for example due to an inhomogeneous magnetic field $B_{\rm dc}$.
A further reason may be a dependence of the g-factors on the
particular quantum dot, or a difference in the hyperfine interaction
\cite{jouralev,inarrea}.  Then, the states $|S_0\rangle$ and
$|T_0\rangle$ mix and, thus, only the triplet states $|+\rangle$ and
$|-\rangle$ suffer from spin blockade.  Moreover, since for an
inhomogeneous Zeeman splitting, $B_{\rm ac}$ can only be at
resonance with the electrons in one of the quantum dots, the
trapping in the triplet subspace is lifted and a finite current can
flow; see figure~\ref{ifrBac}.

The contribution of these two effects---single electron spin
resonance and the suspension of channel blocking by
singlet-triplet mixing---results in a non-monotonic dependence of
the Fano factor, which is considerably enhanced in the vicinity of
the degenerate Zeeman splitting $\Delta_{\rm L}\approx\Delta_{\rm
R}$.  In that region, the rotation of the two electrons still
dominates and increases the blocking time which is much longer
than the time lapses with finite conduction.  This leads to an
enhanced electron bunching which is manifest in the larger Fano factor
shown in figure~\ref{ifrBac}. This super-Poissonian behaviour is
accompanied by an enhanced spin correlation coefficient $r$.
Once single-electron spin resonance dominates owing to large Zeeman
inhomogeneities, the Fano factors of both the charge current and the
spin current saturate at the values $F=3$ and $F_\sigma=2$,
respectively, while the spin correlation coefficient becomes $r=1/2$.

\subsection{Crossover to the incoherent regime: unblocking by relaxation}

The presence of inelastic spin scattering processes damps the
coherent rotations within the triplet subspace resulting in a
finite population of the singlet subspace.  Then, a finite current
emerges. We treat relaxation processes phenomenologically by
introducing in the master equation spin relaxation and spin
dephasing times, $T_1$ and $T_2$, respectively.

\begin{figure}[t]
\begin{center}
\includegraphics[width=2.95in,clip]{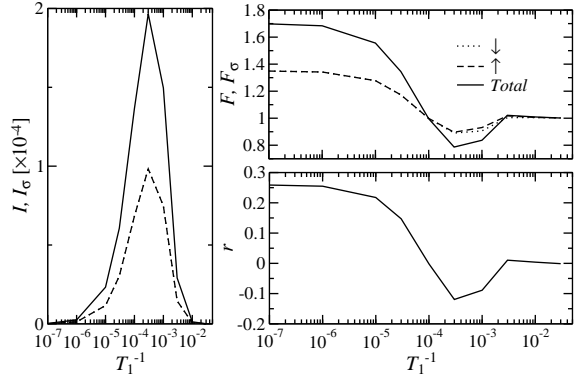}
\end{center}
\caption{\label{ifrRelBac} Current (in $\mathrm{mV}/\hbar$), Fano factors and spin
correlation as a function of the spin relaxation rate. We have
considered $T_2=0.1T_1$ and the same parameters as in
figure~\ref{ifrBac}, except for $\Delta_{\rm L}=\Delta_{\rm R}=0.026$. }
\end{figure}

The results for the Fano factors and the spin correlation for
$\Delta_{\rm L}=\Delta_{\rm R}$ shown in figure~\ref{ifrRelBac}
indicate the existence of two different regimes: When spin
relaxation processes represent merely a perturbation, only a small
leakage current flows and the Fano factors approach $F=5/3$ and
$F_\sigma=4/3$, while the spin correlation coefficient becomes
$r=1/4$. As spin relaxation becomes more important, the current
increases to a maximum which marks the crossover to the incoherent
regime.  At the maximum, the current and the spin current are
sub-Poissonian, $F,F_\sigma<1$.  For even faster relaxation
processes, typically for $T_1<2\pi/\Omega_{\rm ac}$, both spin
rotation and inter-dot tunnelling---which are considered to be of the
same order here, with Rabi frequencies $\Omega_{\rm ac}=2B_{\rm ac}$
and $\Omega_{\rm LR}=2\sqrt{2}t_{\rm LR}$\cite{esr}---are no
longer effective, so that spin relaxation takes place immediately
after spin rotation. Consequently, no electron flow is observed
\cite{esr}. As expected, in this regime the currents are almost
Poissonian and the spin projections are uncorrelated.

\section{Pumping and photon-assisted transport}

\begin{figure}[t]
\begin{center}
\includegraphics[width=2.95in,clip]{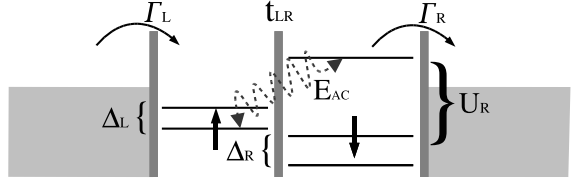}
\end{center}
\caption{\label{pump} Pumping configuration, where spin blockade
is removed by photon assisted processes through the contact barriers.
}
\end{figure}

Not only an ac magnetic field, but also an ac electric field can
support the electron transport by exciting electrons to states
which lie above the chemical potentials of both leads and, thus,
are energetically forbidden. Then, the triplet states have finite
lifetimes due to photon absorption processes so that spin blockade
is resolved \cite{pssb,hansonrev}.
A paradigmatic setup for which resonant ac electric fields can
lead to finite charge \cite{staffordWingreen} or spin currents
\cite{spinpump} even in the absence of a source-drain voltage
($\mu_{\rm L}=\mu_{\rm R}=\mu$) is the spatially asymmetric
\textit{pump} configuration of the double dots sketched in
figure~\ref{pump}.  There the asymmetry is provided by the different
intra-dot interactions $U_{\rm L}>U_{\rm R}$.  Pumping occurs when the
frequency of the electric ac field provides the energy
necessary to transfer the electron from the left quantum dot to an
energetically higher singlet state $|S_{\rm R}\rangle$ in the right
dot.  For this process, the resonance condition reads
$\hbar\omega=\hbar\omega_0=\varepsilon_{\rm R}-\varepsilon_{\rm L}+U_{\rm
R}-U_{\rm LR}$.  As before, the singly occupied states of both dots
are below both chemical potentials, $\varepsilon_l<\mu$, while the
doubly occupied singlets lie well above, $U_l+\varepsilon_l>\mu$.
We restrict ourselves to the pumping configuration, because increasing the
bias voltage reduces the contribution of photon assisted processes
through the contact barriers\cite{Stoof,leganes} which are essential
for the removing spin blockade. 

The ac electric field is modelled as an oscillatory dipole
potential, i.e.\ as a time-dependent energy shift with a phase
$\pi$ between the two dots.  Then the Hamiltonian (\ref{h0})
acquires a term
\begin{equation}
\hat H_{\rm ac}(t)
=\frac{eV_{\rm ac}}{2}\left(\hat n_{\rm L}-\hat n_{\rm R}\right)
\cos(\omega t).
\end{equation}
By the unitary transformation $\hat{U}(t)={\rm e}^{\rmi(eV_{\rm
ac}/2\hbar\omega)(\hat{n}_{\rm L}-\hat{n}_{\rm R})\sin\omega t}$, the
time dependence is transferred to the tunnel couplings according to
\begin{equation}
\hat{H'}(t)=\hat
U(t)\left(\hat H-\rmi\hbar\partial_t\right)\hat U^\dagger(t)=\hat
H_{\rm 0}+\hat{H'}_{\rm LR}(t)+\hat{H'}_{\rm T}(t)+\hat H_{\rm
leads},
\end{equation}
where
\begin{eqnarray}
\label{hopp}
\label{hoppingTrans}
\hat{H'}_{\rm LR}(t)
&=& \sum_{\nu=-\infty}^\infty (-1)^\nu J_\nu\left(\frac{eV_{\rm
    ac}}{\hbar\omega}\right)\sum_{\sigma}\left(t_{\rm LR}
    e^{i\nu\omega t}\hat{c}_{\rm L\sigma}^\dagger\hat{c}_{\rm
    R\sigma}+{\rm H.c.}\right),
\\
\hat{H'}_{\rm T}(t)
&=& \sum_{\nu=-\infty}^\infty(-1)^\nu
    J_\nu\left(\frac{eV_{\rm ac}}{2\hbar\omega}\right)\sum_{lk\sigma}
    \left(\gamma_{l}e^{i\nu\omega
    t}\hat{d}_{lk\sigma}^\dagger\hat{c}_{l\sigma}+{\rm H.c.}\right).
    \label{HTtransform}
\end{eqnarray}
Thus the driving effectively renormalises the inter-dot tunnelling by
the Bessel function $J_\nu$, where the index $\nu$ reflects the
number of photons involved. This creates dynamical charge
localisation under conditions that we specify below.  The
corresponding photon-assisted tunnelling rates turn out to be
\begin{eqnarray}
\label{Gammasac}
\Gamma_{mn}
&=& \frac{2\pi}{\hbar}\sum_{l\nu}d_lJ_\nu^2\left(\frac{eV_{\rm ac}}{2\hbar\omega}\right)
    |\gamma_l|^2\left(1-f_l(\hbar\omega_{mn}+\nu\hbar\omega)\right),
\\
\Gamma_{mn}
&=& \frac{2\pi}{\hbar}\sum_{l\nu}d_lJ_\nu^2\left(\frac{eV_{\rm ac}}{2\hbar\omega}\right)
    |\gamma_l|^2f_l(\hbar\omega_{mn}+\nu\hbar\omega),
\end{eqnarray}
where the former rate governs processes $|n\rangle\to|m\rangle$ that
remove an electron from the double dot, while the latter refers to
adding an electron to the system. The density of states in the leads,
$d_l$, is assumed to be constant. 

If a magnetic field is applied such that the energies of the spin-down
electrons are shifted by $\Delta_l$, the spin-up electron is
delocalised within the double quantum dot occupying the states
$|{\uparrow},{\downarrow}\rangle$ and
$|0,{\uparrow}{\downarrow}\rangle$ until one of the electrons tunnels
to the right lead. If the Zeeman splitting is the same in both dots,
the spin-down electron can at the same time be delocalised between
$|0,S_{\rm R}\rangle$ and $|{\downarrow},{\uparrow}\rangle$ for the
same frequency.

\begin{figure}[t]
\begin{center}
\includegraphics[width=2.95in,clip]{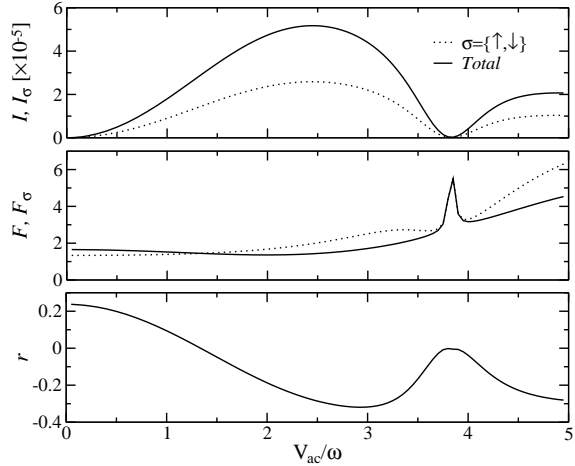}
\end{center}
\caption{\label{ifrEac} Current (in $\mathrm{mV}/\hbar$), Fano factors
and spin correlation as a function of the electric field intensity for
$\omega=\omega_0$, $t_{\rm LR}=0.005$, $\varepsilon_{\rm L}=0.4$,
$\varepsilon_{\rm R}=0.25$, $U_{\rm R}=1$, $U_{\rm LR}=0.5$,
$\Delta_{\rm L}=\Delta_{\rm R}=0.026$, $\mu_{\rm L}=\mu_{\rm R}=1.2$,
$\Gamma_{\rm L}=\Gamma_{\rm R}=10^{-3}$ (energies in meV).}
\end{figure}

At low ac intensities, photon-assisted tunnelling is not
significant. Indeed, using the approximation $J_\nu(x) \approx
\delta_{\nu,0}$ for small $x$, we find that photon-assisted
tunnelling rates become identical to the rates for the static
system discussed in section~\ref{undriven}; cf.\ figure~\ref{ifrEac}.
When increasing the intensity, both the current and the noise
exhibit a more involved behaviour stemming from the amplitude and
frequency dependence of the tunnel rates via the arguments of the
Bessel functions $J_1(eV_{\rm ac}/\hbar\omega)$ and $J_1(eV_{\rm
ac}/2\hbar\omega)$ which govern inter-dot tunnelling and photon-assisted
tunnelling, respectively. When the ac intensity approaches the
value for which $J_1(eV_{\rm ac}/\hbar\omega)$ assumes its maximum,
inter-dot tunnelling is most effective and, thus, the current
assumes its maximum as well, while the Fano factor assumes a
minimum \cite{strass}. The spin Fano factor, by contrast, does not
assume a minimum and, consequently, the spin correlation $r$
becomes negative. Increasing the ac intensity further reduces the
net current because in this limit, the rates of left-to-right
tunnelling approach those of the opposite processes.  This
cancellation does not affect the noise strength $S$, so that a smaller
current leads to an increasing Fano factor.

Particularly interesting is the condition $J_1(eV_{\rm
ac}/\hbar\omega)=0$ for which dynamical charge localisation is found.
Then one-photon interdot tunnelling is suppressed due to the
renormalisation factor in the hopping term (\ref{hopp}). Since for
our configuration these processes represent the main contribution
to the transport, the current becomes strongly suppressed, as can
be seen in figure~\ref{ifrEac} in the region close to $eV_{\rm
ac}\approx 3.8\hbar\omega$.  Nevertheless, photon-assisted tunnelling
through the contacts is still noticeable by its
contribution to the current fluctuations, leading to a sharp peak
in the Fano factor \cite{genovaPAT}.  This behaviour is different from
the one found in the large bias limit, where current suppression is
associated with Poissonian noise \cite{camalet}.

Another important factor for spin blockade is the difference between
the Zeeman splittings, as discussed in section \ref{esr}.  If
$\Delta_{\rm L}=\Delta_{\rm R}$, the states $|{\uparrow},
{\downarrow}\rangle$ and $|{\downarrow},{\uparrow}\rangle$ are
indistinguishable, then the inter-dot singlet $|S_0\rangle$ and the
triplet $|T_0\rangle$ influence the dynamics. Since inter-dot
tunnelling does not change the total spin, it can only occur within the
singlet states $|S_0\rangle$ and $|S_{\rm R}\rangle$. Thus, not only
the states $|{\uparrow},{\uparrow}\rangle$ and
$|{\downarrow},{\downarrow}\rangle$ contribute to the transport
blocking, but also the triplet state $|T_0\rangle$.

\begin{figure}[t]
\begin{center}
\includegraphics[width=2.95in,clip]{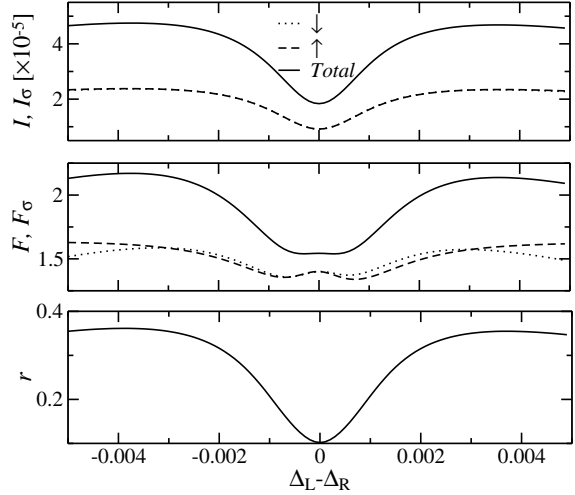}
\end{center}
\caption{\label{ifrEacDeltas} Current, Fano factor and spin
correlation as functions of the Zeeman inhomogeneity (by varying
$\Delta_{\rm L}$) for $eV_{\rm ac}=\hbar\omega=\hbar\omega_0$ and the same
parameters as in figure~\ref{ifrEac}. Though the spin currents almost
coincide, as seen in the upper panel, their dynamics can be
distinguished in the noise components.}
\end{figure}

On the other hand, if the Zeeman splittings are different,
$\Delta_{\rm L}\ne\Delta_{\rm R}$, spin blockade is less effective due
to the mixing between the states $|T_0\rangle$ and $|S_0\rangle$
\cite{esr}. Then the current increases, as can be seen in
figure~\ref{ifrEacDeltas}. The reduced the number of blocking states
again diminishes the Fano factor. If the difference between the Zeeman
splittings is large enough, we find resonant one-photon inter-dot
tunnelling for electrons with one particular spin polarisation---here
the spin-up polarisation.  Consequently, the transport becomes spin
dependent, as can be appreciated in the different Fano factors in
figure~\ref{ifrEacDeltas}.

\section{Conclusions}

In this work we have studied the suspension of spin blockade in the
electron transport through coherently coupled double quantum dots by
ac driving fields and also by thermal excitations.  In particular, we
focussed on the associated shot noise characterised by the
Fano factors for the electron and the spin currents, as well as on
the spin-spin correlation coefficient.

One way to resolve spin blockade is the application of magnetic fields
that cause Zeeman splitting and spin rotation.  Most interestingly, we
found that the current noise for both charge transport and spin
transport depends sensitively on whether the Zeeman splitting is the
same or different for the two dots.  For degenerate Zeeman splitting,
the eigenstates are delocalised and the transport across the whole
sample is dominated by tunnel events which obey Poissonian statistics.
Thus the Fano factors tend to be close to unity.  If the
degeneracy is absent, transport is dominated by lapses of time at
which spin blockade is lifted.  During these lapses, we observe the
transport of bunches of electrons with correlated spins.

Considering also spin relaxation, we find that such incoherent
processes can even contribute to the suspension of spin blockade:
With an increasing relaxation rate, transport is enhanced
and becomes more regular and even sub-Poissonian.  When the
predominantly incoherent regime is entered, the Fano factor increases
again until we eventually reach an incoherent regime with
Poissonian tunnel currents.

Time-dependent fields can also resonantly excite electrons to orbitals
with energies above both Fermi surfaces.  In that way, they may cause
photon-assisted transport and, in asymmetric configurations, electron
pumping.  The latter results in a contribution to the dc current at zero bias voltage.
Photon assisted transport can be described by performing a unitary
transformation of the Hamiltonian such that only the tunnelling matrix
elements are explicitly time dependent. This renormalises the
tunnel couplings such that they depend on driving amplitude and
frequency.  The main contribution to interdot hopping
can even vanish, so that for specific driving parameters, the
electrons become localised within one dot.  Consequently, the current
is almost suppressed.  Nevertheless, the current fluctuations may stay
at a significant level and, accordingly, the shot noise level is
super-Poissonian.  With an additional Zeeman splitting, the resonance
condition becomes spin dependent. Then photon-assisted transport
favours one particular spin projection and the two spin currents are no
longer identical.  This also affects the spin shot noise, although the
difference in the corresponding Fano factors turns out to be
relatively small.

In conclusion, our results underline that ac fields can have
intriguing consequences for spin-dependent transport.  Depending on
the type of excitation, both charge and spin transport can
occur in bunches.  We have shown that this should be clearly visible
in the corresponding measurements of the Fano factor and, most likely,
also in higher orders of the full-counting statistics.  For driven
spin transport this still is awaiting closer investigation.

\ack
Work supported by the M.E.C. of Spain through Grant No.\ MAT2005-00644.
SK acknowledges support by the DFG through SFB 484 and the Excellence
Cluster ``Nanosystems Initiative Munich (NIM)''.


\end{document}